# Strategic Innovation Through Outsourcing – A Theoretical Review

*Marfri Gambal, Aston University*
*Aleksandre Asatiani, University of Gothenburg*
*Julia Kotlarsky, University of Auckland*

## Abstract

Competition in the Information Technology Outsourcing (ITO) and Business Process Outsourcing (BPO) industry is increasingly moving from being motivated by cost savings towards strategic benefits that service providers can offer to their clients. Innovation is one such benefit that is expected nowadays in outsourcing engagements. The rising importance of innovation has been noticed and acknowledged not only in the Information Systems (IS) literature, but also in other management streams such as innovation and strategy. However, to date, these individual strands of research remain largely isolated from each other. Our theoretical review addresses this gap by consolidating and analyzing research on strategic innovation in the ITO and BPO context. The article set includes 95 papers published between 1998 to 2020 in outlets from the IS and related management fields. We craft a four-phase framework that integrates prior insights about (1) the antecedents of the decision to pursue strategic innovation in outsourcing settings; (2) arrangement options that facilitate strategic innovation in outsourcing relationships; (3) the generation of strategic innovations; and (4) realized strategic innovation outcomes, as assessed in the literature. We find that the research landscape to date is skewed, with many studies focusing on the first two phases. The last two phases remain relatively uncharted. We also discuss how innovation-oriented outsourcing insights compare with established research on cost-oriented outsourcing engagements. Finally, we offer directions for future research.

**Keywords**

Business process outsourcing; BPO; information technology outsourcing; ITO; literature review; outsourcing; strategic innovation; theoretical review

# Introduction

Over the last decade, competition in the Information Technology Outsourcing (ITO) and Business Process Outsourcing (BPO) sector has shifted from transactional engagements, towards trust-based partnerships (Susarla and Mukhopadhyay, 2019). In such partnerships, there is growing emphasis on *strategic innovation* (Weeks and Feeny, 2008). Strategic innovations substantially enhance the client's overall competitiveness (Lacity and Willcocks, 2013), contribute to its strategic objectives (Oshri et al., 2015; Su et al., 2015), and support its business-wide transformation programs (Dibbern and Hirschheim, 2020; Asatiani et al., 2019; Langer and Mani, 2018).

While there is ample evidence of clients expecting providers to deliver innovation (Oshri et al., 2018, 2015; Su et al., 2015; Susarla and Mukhopadhyay, 2019), recent studies also expose potential tensions, especially between cost-oriented and innovation-oriented engagements (Aubert et al., 2015; Kotlarsky et al., 2016). Cost-oriented engagements are predominantly anchored in a transactional mindset wherein clients and providers exchange a service for a fee (Aubert et al., 2015). They build on expectations and sanctions that are formally defined ex ante (Susarla and Mukhopadhyay, 2019). Providers offer specialized competences that allow clients to realize cost savings and improve their operational efficiency (Levina and Ross, 2003; Nevo and Kotlarsky, 2014).

Cost-oriented engagements however fail to accommodate core features associated with strategic innovation, which calls for a creative mindset (Gurteen, 1998) and requires slack resources with a risky return model (Garcia-Granero et al., 2015). The result is notable outsourcing management quandaries. One example is the highly detailed Service Level Agreements (SLAs), which while generally indispensable to ensuring service quality consistency in cost-oriented outsourcing contracts, discourage experimentation and risk-taking on the part of the provider (Aubert et al., 2015).

Driven by our interest in understanding how clients and providers can deal with these challenges, we turned to the wider management literature, identifying a range of studies outside mainstream IS journals that examine links between innovation and IS outsourcing. For example, in the innovation management literature, Roy and Sivakumar (2012) note that outsourcing engagements can yield radical innovations, while in the strategy literature, Chatterjee (2017) reveals how innovative provider solutions are tailored to meet the business objectives of individual clients.

While research on innovation and outsourcing is evidently flourishing, the landscape to date remains fragmented. This motivated us to engage with the wider body of business management literature to bring currently disconnected insights from related studies, spread across multiple management fields, into the IS outsourcing domain. We conducted a review that largely reflects the main principles of Paré et al.'s (2015) approach to theoretical review. The resulting article set comprised 95 studies published between 1998 and 2020 from multiple research streams, among which IS, innovation, and general management have the strongest presence. Our subsequent analysis followed Wolfswinkel et al.'s (2013) grounded theory techniques adapted for literature reviews.

This review offers three major contributions to the IS outsourcing literature. First, we consolidate a large body of knowledge into an integrative framework. Second, we advance discussion on how innovation-oriented outsourcing research insights compare with cost-oriented outsourcing. Our third contribution is five future research directions that build on our integrative framework.

# Background

The scholarly perception of what is considered "innovation" in the outsourcing context has changed as the ITO and BPO industries have evolved and matured. Earlier studies on ITO from the early-to-mid 1990s view the mere decision by a firm to outsource some or all of its IT functions to an external provider as an innovation, (Grover et al., 1994a; Gurbaxani, 1996; Loh and Venkatraman, 1992a, 1992b; Venkatraman et al., 1994). Papers by leading IS researchers, such as Loh and Venkatraman (1992a), Ang and Cummings (1997) and Hu et al. (1997), report on empirical studies that use theories of innovation adoption and diffusion to model the acceptance and spread of ITO itself.

In the late 1990s, the urgent need for companies to prepare their systems for the new millennium (the rollover from the year 1999 into 2000 – commonly referred to as the Year 2000 or Y2K problem) led to a significant expansion of the ITO industry, which has been growing ever since. Carmel and Agarwal (2002) capture the maturation of offshore ITO in moving away from a focus on cost savings towards a proactive strategic focus.

A decade later, as the boundaries between ITO and BPO were becoming increasingly blurred (Lacity et al., 2016), competitive momentum in the outsourcing industry started shifting towards a value proposition that includes innovative solutions with a business-wide impact on top of cost savings and freeing up resources for core activities. In this study, we focus on this strategic aspect of innovation, which has become one of the main trends in the outsourcing industry, attracting significant attention from IS scholars working on outsourcing-related topics.

## What is strategic innovation through outsourcing?

Interest in understanding how innovation can be delivered in the outsourcing context is growing (Aubert et al., 2015; Oshri et al., 2018). Weeks and Feeny (2008) offer a refined categorization of innovation specifically emerging from the outsourcing context. It distinguishes between operational innovation, business process innovation, and strategic innovation. Strategic innovation, defined as ways to "significantly enhance the firm's product or service offerings for existing target customers, or enable the firm to enter new markets" (Weeks and Feeny, 2008), tends to be challenging for firms to achieve (Oshri et al., 2015; Weeks and Feeny, 2008). Weeks and Feeny's (2008) definition of strategic innovation reflects the radical/exploratory concept of innovation discussed in the innovation and strategy literature. Such innovations help firms offer new products and/or service lines (Droege et al., 2009), facilitate new market entries (Berry et al., 2006), or introduce new distribution channels (Jansen et al., 2006).

Strategic innovation in an outsourcing context tends to emerge in ongoing engagements; that is, after an outsourcing contract is awarded and a relationship between the client and supplier develops (Aubert et al., 2015; Oshri et al., 2018, 2015; Su et al., 2015; Weeks and Feeny, 2008). As illustrated in Weeks and Feeny's (2008) study, clients tend to initially outsource for cost savings, then gradually shift their attention to quality, and then to innovation as the outsourcing relationships matures.

# Method

## Literature search and selection process

Our search started with scoping out the state of the research landscape in an unstructured fashion to gain initial understanding and identify seminal works. We noticed that most relevant works are published in outlets listed in four subject categories of the Chartered Association of Business Schools' (CABS) *Academic Journal Guide*: information management, innovation, general management, and strategy. We then created a preliminary list of 3, 4 and 4*-rated journals (shown in Appendix I) from these four CABS subject categories to be used in the structured search subsequently conducted, which consisted of three steps.

***Step 1:*** We turned to the publisher database of each journal, using the terms "innovation" AND "outsourcing" in title, abstract or keyword searches to locate relevant articles published between 1998 and 2020. If the search engine of a journal's publisher database only featured limited search options, we additionally drew on EBSCO Business Source Premier, ProQuest, or JSTOR databases, depending on their embargo periods for the specific journal. We retrieved 133 papers (see database search results based on the preliminary list of journals in Appendix I).

***Step 2:*** We applied quality-based, content-based, and time-based inclusion and exclusion criteria (see Table 1 and further details in Appendix II). Removing papers that did not meet our inclusion criteria reduced the sample to 39 papers.

***Step 3:*** We ran three rounds of forward and backward citation searches, starting with the 39 selected papers, and continuing the process for articles added after each round. The sample thereby increased to 95 papers. Citation searches for papers added after the third round did not yield any new relevant papers. A summary of our final journal and article set is included in Appendix III.

Interestingly, of final article sample comprising 95 papers (see Appendix IV), only 40 are published in mainstream IS journals. We view this as a notable indicator of the need for the integrative perspective offered in this review.

**Table 1: Inclusion and exclusion criteria**

| **Criterion type** | **Description** |
|---|---|
| Quality-based | Only include papers published in peer-reviewed journal from 3, 4, and 4*-rated journals listed in the CABS Academic Journal Guide 2018.<br>*Excluded:*<br>*Papers published in lower-rated or non-listed outlets, and any other type of publication, such as books, book reviews, conference papers, teaching cases or industry reports.* |
| Content-based | Only include papers that discuss innovation in the context of ITO and BPO engagements.<br>*Excluded:*<br>*Research that discusses innovation in other outsourcing contexts, such as contract manufacturing* (e.g., Dabhilkar et al., 2009; Triguero and Córcoles, 2013)*, R&D outsourcing and crowdsourcing.* |
| | Only include papers that discuss innovations featuring at least one of two properties, while not contradicting the other:<br>1. High degree of uncertainty associated with innovation outcomes (i.e., final product/service not known *a priori).*<br>2. Final outputs materialize in the form of complex, IT-enabled products and services.<br>3. Outcome is of strategic importance to the client (i.e., impacts important areas (if not the entire business) of the client and improves the firm's overall competitiveness).<br>*Excluded:*<br>*Journal articles that discuss other innovation concepts such as the outsourcing decision as an innovation* (e.g., Hu et al., 1997; Loh and Venkatraman, 1992) and *physical, non-IT product innovations* (e.g., Marion and Friar, 2012; Mikkola, 2003; Park et al., 2018; Takeishi, 2002). |
| Time-based | Search limited to articles published between 1998 and 2020. |

### Analyzing the final article sample

Our analysis was guided by Wolfswinkel et al.'s (2013) grounded theory techniques for thematic analysis based on an iterative process of 'open coding', 'axial coding', and 'selective coding'. We organized our codes into first-order concepts, second-order themes, and aggregate dimensions, as suggested by Gioia et al. (2013). Our coding structure is presented in Appendix V. We started the process with open coding, which resulted in several first-order concepts. A list of these concepts with key references is included in Appendix VI. We then proceeded with axial coding to categorize first-order concepts into second-order themes representing distinct aspects of strategic innovation in the outsourcing context. Our second-order themes are briefly outlined in Appendix VII. Lastly, we performed selective coding, which resulted in four aggregate dimensions – Antecedents, Arrangements, Generation and Outcomes. In the next section we report our findings, discussing each dimension and its second-order themes in greater detail.

## Theoretical review of strategic innovation through outsourcing

### Strategic innovation through outsourcing: research landscape

We distinguish between the three scenarios discussed in the literature in relation to strategic innovation through outsourcing; we term them "pre-contract", "during-contract" and "post-contract" to reflect the temporal dimensions of an outsourcing engagement.

The *pre-contract* scenario captures studies dealing with the complexities and contrarieties of the decision to leverage outsourcing for strategic innovation. This decision can prove challenging for first-generation outsourcing clients in particular. The *during-contract* scenario captures relevant literature on existing outsourcing engagements. Here, the client is already contractually tied to one or several providers. This literature addresses emerging innovation opportunities, and associated challenges with pursuing such opportunities. In the *post-contract* scenario, scholarly evaluations of realizable strategic innovation-enabled business outcomes take center stage.

Additionally, four innovation-centric phases emerged from our analysis of the literature – *antecedents* of the strategic innovation decision, engagement *arrangements*, strategic innovation *generation*, and related *outcomes*. They broadly outline the process of achieving strategic innovation, as suggested in the reviewed literature, and illustrated in our integrative framework (Figure 1). To situate these phases in an outsourcing context, the three outsourcing scenarios (pre-contract, during-contract, and post-contract) have been woven into the framework. Each phase includes second-order themes, shown as bullet points in the framework.

In the framework, some innovation-centric phases are depicted stretching across multiple outsourcing scenarios to underscore that firms can be in different outsourcing engagement stages but face similar challenges when pursuing strategic innovation. For instance, concerning the arrangement theme, clients in a pre-contract scenario need to set up the outsourcing engagement from the ground up, choosing between one or multiple providers, pricing models, and contract completeness. Clients in a during-contract scenario, in contrast, may already have such structures in place, but need to modify them to facilitate the achievement of strategic innovations.

Next, we present our findings relating to each phase. For each theme, we first introduce an overview of commonly applied theoretical perspectives and subsequently link these to prior research insights.

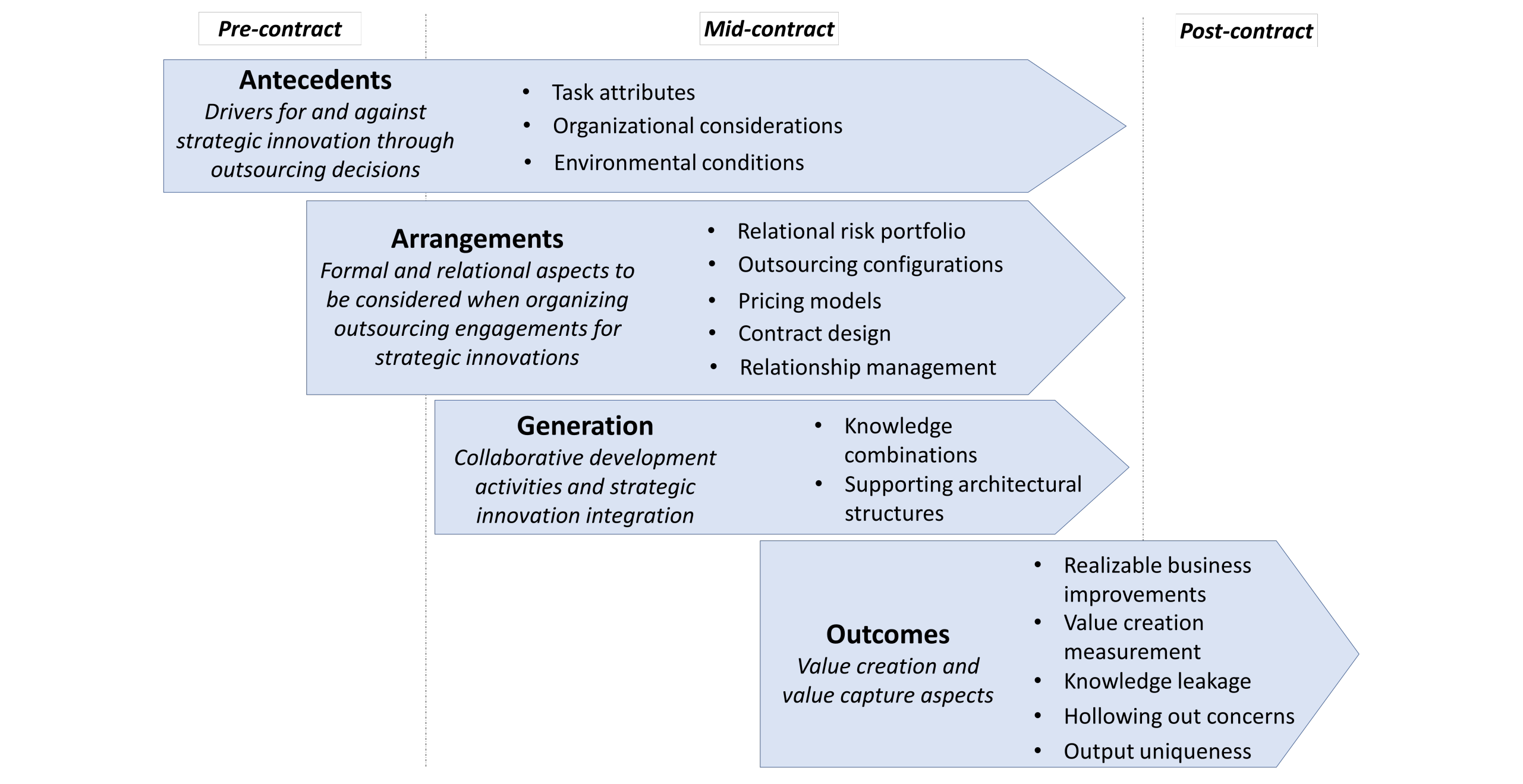

**Figure 1. Strategic innovation through outsourcing: Integrative framework**

## Phase 1: Antecedents of the strategic innovation through outsourcing decision

The literature examines the various motivations driving the decision to pursue strategic innovation in outsourcing. We divided these into three themes: task attributes, organizational considerations, and environmental conditions (see Figure 2).

**Antecedents**
*Drivers for and against the decision to pursue strategic innovation in outsourcing*

| Main themes | Characteristics specific to outsourcing | Key theoretical perspectives |
|---|---|---|
| Task attributes | Exploratory tasks span across organizational boundaries | TCE, KBV |
| Organizational considerations | Access to specialized external IT resources (client view) / deep domain knowledge (provider view) | Relational view, resource theories, evolutionary perspectives, organizational design |
| Environmental conditions | IT-based services commoditization pressures providers to differentiate, e.g., by engaging in innovation efforts | Network perspectives (client), internationalization perspectives (provider) |

**Figure 2. Notable characteristics and theoretical perspectives in the "Antecedents" phase**

### *Task attributes*

The literature generally draws on two reference theories to examine innovation task attributes and their compatibility with outsourcing: transaction cost economics (TCE) (Williamson, 1985, 1975) and the knowledge-based view (KBV) (Kogut and Zander, 1992).

A main premise of TCE asserts that transactions involving highly specific assets and high uncertainty incur lower transaction costs within the firm and should be internalized (Aubert et al., 2004; Walker and Weber, 1984). Prior research associates the strategic innovation task environment with high asset specificity and high uncertainty. High asset specificity stems from the need to build deep relationship-specific knowledge (Oshri et al., 2018; Weeks and Feeny, 2008) and high uncertainty from the difficulty of defining the exact nature of activities in advance because the final product is not known *a priori* (Miranda and Kavan, 2005; Oshri et al., 2015). Accordingly, some studies indicate that innovation can be better performed internally (Qu et al., 2010; Weigelt and Sarkar, 2012).

Applying the KBV lens has led to similar conclusions by building on the notion that firms function as social communities. According to this group of studies, the knowledge held by organizational members can be better combined to create innovations than in the market because communities feature a common language and organizing principles (Bunyaratavej et al., 2011; Takeishi, 2002; Verwaal, 2017). In this vein, Qu et al. (2010) and Zimmermann et al. (2018) respectively show that internal IT departments outperform outsourcing engagements in knowledge sharing and coordination and the extent of knowledge sharing is greater in captive sourcing arrangements than in external sourcing modes.

Overall, the rationale behind the decision to pursue strategic innovation in outsourcing engagements does not seem to originate from transaction cost and knowledge sharing efficiency considerations. Evidently, uncertain, difficult to codify, and complex tasks involving extensive relationship-specific knowledge can be carried out more optimally in vertically integrated models. However, prior research has uncovered encouraging drivers that move beyond a focus on task attributes.

### *Organizational considerations*

We identified four prominent organizational-level drivers variously linked with four co-existing theoretical approaches that support the strategic innovation through outsourcing decision. They are: 1) access to specialized resources, from the resource-based view (RBV) (Barney, 1991) and core competency perspectives (Prahalad and Hamel, 1990); 2) compatible business strategies, based on organizational design perspectives; 3) evolving demands for more value, based on evolutionary perspectives; and 4) the outsourcing relationship as an enabler of competitive advantage, based on the relational view (Dyer and Singh, 1998).

Gaining access to specialized resource pools is a compelling driver for strategic innovation through outsourcing. For clients, outsourcing presents a practical alternative through substituting weak internal IT resources with those of a best-in-class provider (Shi, 2007). It is the potent combination of providers' advanced IT resources and the client's accumulated cross-industry domain knowledge that can lead to strategic innovations (Kedia and Lahiri, 2007). For providers, the domain-specific knowledge of clients is valuable (Arora et al., 2001). These intellectual resources are hard to learn, systematize or replicate for others, but are critical to creating superior value for clients and can be effectively cultivated by engaging in joint innovation initiatives (Chatterjee, 2017; Desyllas et al., 2018).

Prior research frequently extends resource-related arguments to organizational design perspectives focused on the interplay between resource stocks and business strategies (Mukherjee et al., 2013). Strategic innovation appears to only rarely be a standalone outsourcing project objective; it is more often tied to a client's business goals (Jensen, 2009), including technical leadership (Choudhury and Sabherwal, 2003; Gozman and Willcocks, 2019; Kern et al., 2002) and business transformation (Kedia and Lahiri, 2007; Linder, 2004; Mani et al., 2010). Conversely, providers are more inclined to engage in strategic innovation through outsourcing efforts when they follow a differentiation-based business strategy focused on novel resource combinations to deliver custom solutions that support the client's business objectives (Desyllas et al., 2018).

Evolutionary perspectives principally suggest that organizations eventually shift to innovation objectives after experiencing satisfactory results with operational services. A prominent conceptual learning curve framework by Rottman and Lacity's (2006) is applied by Weeks and Feeny (2008) to divide outsourcing relationships into four stages. After a client learns about the potential benefits of outsourcing (first stage), it gradually moves from cost (second stage) to quality (third stage) and then to innovation objectives (fourth stage).

Lastly, the relational view (Dyer and Singh, 1998) posits that business relationships yield competitive advantages for the involved parties (Miranda and Kavan, 2005) via relationship-specific assets, knowledge-sharing routines, and complementary resources. Relationship-specific assets take the form of synergistic knowledge bases cultivated by the parties for strategic innovations (Weeks and Feeny, 2008). Knowledge-sharing routines are then necessary to effectively leverage these knowledge bases (Dyer and Singh, 1998). These routines are a function of prior related knowledge, or absorptive capacity as described by Cohen and Levinthal (1990). Bilateral absorption effects have been noted, wherein the client absorbs its provider's technological knowledge, while the provider's capacity progressively expands with client-specific knowledge (Oshri et al., 2018; Weeks and Feeny, 2008). In practice, complementary resources may take the shape of jointly owned value-creation centers (Kotlarsky et al., 2016).

### *Environmental conditions*

Some studies report notable industry-level drivers for the decision to pursue strategic innovations through outsourcing. For clients, the increasing shift of competition from between firms to between networks is a major factor. For providers, increasing commoditization of business services is eroding their competitiveness based on labor arbitrage alone.

A value network perspective provides a theoretical basis for examining the advent of strategic innovation through outsourcing from the client perspective (Manning et al., 2018; Van de Ven, 2005). Building on the open innovation paradigm, it emphasizes the new business opportunities that become accessible through joining forces (Aubert et al., 2015). Studies applying this perspective argue the knowledge needed to develop strategic innovations is rarely contained within a single firm, rendering vertical integration unfeasible (Manning, 2013; Van de Ven, 2005). This indicates the shift in competition from between individual firms to between value networks (Van de Ven, 2005). Studies show clients across various industries integrating both established and disruptive providers as nodes in their value networks (Su et al., 2015; Su and Levina, 2011).

Studies have also analyzed the provider industry landscape, detecting cluster developments in popular outsourcing destinations, especially India (Lema et al., 2015; Manning, 2013; Massini and Miozzo, 2012). Such developments are due to the commoditization of knowledge work and growing demand for high-skilled, yet lower cost talent (Manning, 2013). Largely due to the explicit nature of technological knowledge (Chatterjee, 2017; Gopal and Gosain, 2010), this commoditization has increased competition among providers (Arora et al., 2001; Davenport, 2005; Manning et al., 2018).

To stay strategically relevant, they are pressed to craft customized solutions that facilitate the core business activities of their clients (Arora et al., 2001). High demand for the limited supply of highly skilled labor increases wage pressures, in turn indicating that current competitive strategies relying solely on cost advantages may not be sustainable (Arora et al., 2001; Manning, 2013).

#### *Antecedents phase: Conclusion*

The decision to engage in strategic innovation through outsourcing appears to be driven by long-term business development, rather than short-term profitability motives. Examining the task environment in isolation of wider business imperatives suggests the coordination advantages of vertically integrated organizations work against engagement in innovation via outsourcing. However, scholars also recognize the intellectual resources for innovation are increasingly distributed outside the firm's organizational boundaries. Outsourcing practices today offer the potential to pool forces and jointly generate strategic innovations that enable mutually beneficial business outcomes. Within the wider business environment, clients are increasingly competing based on their network. Clients may therefore see their providers as valuable network nodes, while providers need to differentiate themselves in a maturing outsourcing services industry. Table 3 provides a summary of these findings.

**Table 3: Overview of key findings related to the "Antecedents" phase**

| Main theme | Key findings |
|---|---|
| Task attributes | Within-firm theoretical perspectives suggest that uncertain, difficult to codify and complex tasks that extensively involve client firm domain knowledge are better completed internally than via the market. Some conceptual and empirical studies support these suggestions. |
| Organizational considerations | Competitive considerations at an organizational level provide compelling arguments for the strategic innovation through outsourcing decision. They include the development of idiosyncratic, relationship-specific assets, access to specialized resource pools, and enabling business strategies. |
| Environmental conditions | Scholars increasingly highlight emerging forms of competitive value networks from a client firm perspective, and a maturing outsourcing industry from a service provider perspective. |

### Phase 2: Arranging for strategic innovation in outsourcing

Outsourcing arrangements that allow for or facilitate strategic innovations have been widely researched, with studies conducted on relational risks, outsourcing configurations, pricing models, contract design, and relationship management (see Figure 3).

**Arrangements**
*Formal and relational aspects to be considered when organising outsourcing engagements for strategic innovations*

| Main themes | Characteristics specific to outsourcing | Key theoretical perspectives |
|---|---|---|
| Relational risk portfolio | Outsourcing partners are external firms with interests that naturally diverge from those of the client | Agency theory, TCE |
| Outsourcing configurations | Unlike other external sources of innovation, providers are specialized in IT and business services | Supply base management, TCE, resource theories, coopetition |
| Pricing models | Individual remuneration models can be devised to incentivize innovation efforts | Agency theory, contract theory |
| Contract design | The written contract is the formal basis of an outsourcing engagement | Agency theory, TCE, control theory |
| Relationship management | Trustworthiness and commitment need to be fostered after the written contract has been signed | Agency theory, TCE, control theory, relationship theories |

**Figure 3. Notable characteristics and theoretical perspectives in the "Arrangements" phase**

### *Relational risk portfolio*

Outsourcing engagements are known to involve a range of relational risks (Aron et al., 2005; Handley and Benton, 2009; Hoecht and Trott, 2006; Shi, 2007). When leveraged for innovation, these are greatly aggravated by the wider scope for opportunism (Aubert et al., 2015). Typically viewed through the lens of agency theory (Eisenhardt, 1989), such risks primarily stem from information asymmetry and goal incongruence (Langer and Mani, 2018; Roy and Sivakumar, 2012; Wiener et al., 2019). Two types of risks are distinguished herein, namely the adverse selection problem arising from hidden information *ex ante*, and the moral hazard problem arising from hidden action after the contract has been signed (Hart and Holmström, 1987). Other relational risks that recur in prior research are the hold-up problem and knowledge poaching.

Adverse selection problems denote situations wherein the principal (client) does not have complete information about relevant characteristics of the agent (provider) (Roiger, 2006). Consequently, the client is unable to recognize the ideal provider and may select a second-best provider who claims to be best-in-class. The difficulty of conceptualizing innovations in advance complicates realistic appraisals of provider capacity to meet innovation demands (Miozzo et al., 2016).

The moral hazard problem, also referred to as supplier shirking (Handley and Benton, 2009), involves deliberate underperformance by providers while claiming full payment. Drivers for not fulfilling agreed-on responsibilities include self-interest, combined with the imperfect ability of clients to fully observe provider efforts and detect shirking (Aron et al., 2005). In one of their observed cases, Choudhury and Sabherwal (2003) found evidence of a provider shirking testing responsibilities and relying on the client's exhaustive test plans instead.

A key concept in TCE, the hold-up problem (Klein et al., 1978) arises from hidden intentions (Roiger, 2006) in relationship-specific investments that have no value in alternative engagements. In the case of outsourcing, both the client and provider are vulnerable to hold-up. The client may be exposed to providers who intend to deliver only chunks of the developed innovation, or the entire innovation but on terms and at a price that reduce the client's benefits (Aubert et al., 2015). In contrast, the provider may be forced to engage in customizations at rock-bottom prices under threat of the client ending the relationship (Veltri et al., 2008). In both cases, these tangible threats can motivate either victimized party to stop cooperating, or even start retaliating (Frydlinger et al., 2019).

Poaching refers to the provider reselling business-critical knowledge obtained through the relationship to the client firm's competitors (Aron et al., 2005; Clemons and Hitt, 2004; Handley and Benton, 2009). This risk is particularly salient in strategic innovation through outsourcing given the tendency of providers to re-customize solutions for other clients (Desyllas et al., 2018). As Hoecht and Trott (2006) describe, while closer access to client domain knowledge can lead to more useful innovations, the risk of sensitive information leaks increases – creating a challenging trade-off.

### *Outsourcing configurations*

Scholars remain divided as to which outsourcing configuration is most conducive to strategic innovation. Several studies argue for slim supply bases, that is, selectively engaging with one or few providers (Bui et al., 2019; Lee et al., 2004; Su and Levina, 2011; Weeks and Feeny, 2008; Wiener and Saunders, 2014). Conversely, other studies report successful innovation outcomes with broad supply bases (Su et al., 2015). Supply base configuration decisions have notable implications for 1) access to a diversity of ideas, 2) evolution of dependency, and 3) coordination. Few advancements are noted in slim and broad supply base combinations.

Some research suggests that keeping multiple niche players on the radar may result in a continuous flow of diverse ideas for clients (Su et al., 2015), in turn increasing flexibility for accommodating changing requirements (Bui et al., 2019), and allowing the client to better probe the potential of new technologies (Su et al., 2015; Su and Levina, 2011). In contrast, engaging with only one or a few providers narrows the diversity of ideas (Su et al., 2015), but promotes the development of shared language, knowledge, and routines (Bui et al., 2019). These co-developed capabilities in turn facilitate the discovery of business-level innovations more closely suited to the distinctive characteristics of the client (Weeks and Feeny, 2008).

Dependencies develop over time and are a salient issue in slim supply base configurations. Lock-in problems can arise, especially in combination with long-term contracts (Kumar and Snavely, 2004; Su et al., 2015). In the tradition of TCE, switching costs are high when the task environment is characterized by high uncertainty (Cordella and Willcocks, 2012) and the development of highly specific resources that cannot be easily redeployed in different engagements (Lee and Kim, 2010).

When it comes to strategic innovations, this can result in high opportunity costs when clients are locked into a limited set of expertise (Hoecht and Trott, 2006). On the flipside, low dependency may be a continuous reminder that the provider is replaceable (Su et al., 2015), leading to distrust and preempting collaborative problem-solving activities (Miranda and Kavan, 2005).

Broad supply bases require extensive coordination, which translates to high monitoring costs (Su et al., 2015). Moreover, the client may suffer from information losses during communication (Mani et al., 2010), while preventing its providers from having a clear view of all elements associated with the strategic innovation (Aubert et al., 2015). Lastly, multi-sourcing engagements are usually associated with increased competition within the provider portfolio (Bui et al., 2019; Oshri et al., 2019, Wiener and Saunders, 2014).

A promising area only touched on by prior research is configurations leveraging a combination of slim and broad supply bases for strategic innovation. Such combinations are achieved with provider ranking systems. Recent configuration concepts such as the long-tail strategy are based on this idea, wherein a small set of preferred strategic partners are contracted for maintaining platform services (slim supply base), while emerging technologies are leveraged with multiple niche providers on a one-off project basis (broad supply base) (Su et al., 2015).

### *Pricing models*

Prior studies have explored outsourcing engagements based on fixed price contracts (Bui et al., 2019; Mani and Barua, 2015; Miozzo and Grimshaw, 2005; Oshri et al., 2015). Flexible pricing models have also been studied, including time and materials contracts (Bui et al., 2019; Mani and Barua, 2015; Oshri et al., 2015), performance-based contracts (Sumo et al., 2016), and partnership contracts, which share features with joint venture structures (DiRomualdo and Gurbaxani, 1998; Holweg and Pil, 2012; Mani et al., 2010; Oshri et al., 2015).

Fixed price contracts predetermine prices for specified deliverables. Scholars often advise against their stand-alone use for innovation as they require detailed specification of requirements in advance (Bui et al., 2019) and entail high adaptation costs for unforeseeable challenges (Bui et al., 2019; Oshri et al., 2015). Imbalance is also a factor, with the provider bearing the risk of cost escalation, possibly motivating quality cutbacks (Bui et al., 2019).

There is consensus that successful strategic innovation outcomes necessitate flexible pricing models to accommodate the high uncertainty associated with related development tasks. Oshri et al. (2015) find that joint venture contracts, or a joint venture contract in combination with either a fixed price or time and materials contract, amplify the positive effect of relationship quality on the ability to achieve strategic innovation. The success of equity-based contracts in enabling the creation of shared interests and equal sharing of risk and profit is also supported by Holweg and Pil (2012), while Mani et al. (2010) and DiRomualdo and Gurbaxani (1998) emphasize the enhanced facilitation of knowledge transfer. However, equity-based contracts incur considerable set-up costs (Holweg and Pil, 2012).

Performance-based and outcome-based contracts show mixed results. Their potential utility is suggested by DiRomualdo and Gurbaxani (1998), who note "pricing provisions should tie vendor compensation to value received by the client" (p. 10). Sumo et al. (2016) examined two relationships governed by similarly designed performance-based contracts, finding the resulting levels of innovation varied greatly, depending on the client's governance approach during contract execution. Altogether, findings grouped under the pricing model theme show certain contractual pricing models can stimulate strategic innovations. They do however require informal reinforcing conditions, such as high relationship quality (Kedia and Lahiri, 2007; Oshri et al., 2015) or autonomy during contract execution (Sumo et al., 2016).

### *Contract design*

The question of whether more complete contracts are required for strategic innovations, where behaviors and outcomes are extensively formalized, or more incomplete contracts that allow the parties flexibility to deal with new contingencies as they arise (Argyres et al., 2007), has attracted considerable research attention. Prior research on contract completeness is heavily influenced by TCE (Argyres et al., 2007; Goo et al., 2009; Susarla et al., 2010) and control theory (Kirsch, 1997; Wiener et al., 2016).

TCE suggests that economic actors are limited by bounded rationality and can therefore not craft fully complete contracts (Argyres et al., 2007; Susarla et al., 2010). Unable to foresee all possible future contingencies, the actors need to incorporate complex safeguards to protect themselves from hold-up problems arising from contract incompleteness (Susarla et al., 2010). Under control theory, formal safeguards are broadly categorized as outcome controls, such as project milestones, and behavior controls, such as monitoring routines (Kirsch, 1997; Wiener et al., 2016).

Prior insights appear inconsistent regarding contract design. One side suggests that more complete contracts can facilitate innovation. Goo et al. (2009, 2008) show contractual clauses can include an explicit innovation plan specifying the innovation process. A client with knowledge of how innovations can fit with the rest of its organization may be able to formally define related measures (Aubert et al., 2015). Moreover, parties with relationship histories can learn how to specify contractual provisions more effectively over time, enabling them to add more detailed clauses to account for more contingencies (Argyres et al., 2007).

The other side suggests that more complete contracts hinder innovation. Bui et al. (2019) note detailed contracts are a key reason for the lack of strategic innovation. In a similar vein, Langer and Mani (2018) identify incompleteness as an essential feature of well-designed contracts in the case of complex initiatives like innovation involving aspects that are difficult to verify. More complete contracts limit provider flexibility and responsiveness in the face of task or technological changes (Aubert et al., 2015; Miranda and Kavan, 2005). This may even lead to a downward spiral where the provider's inability to innovate drives the client to enforce penalties and monitor the contract more closely (Aubert et al., 2015). Overall, scholars in this camp argue that some best practice contracting principles from traditional outsourcing, like detailing tasks (Holweg and Pil, 2012), can be at odds with making successfully arrangements for strategic innovation (Aubert et al., 2015; Oshri et al., 2018).

### *Relationship management*

Formal obligations to engage in strategic innovation do not guarantee cooperative behavior during the initiative (Lahiri and Kedia, 2009). The outsourcing relationship therefore needs to managed to ensure the client and provider stay committed over the long term (Kedia and Lahiri, 2007). Prior research has applied relationship theories focused on cooperation, interactions, and social and economic exchanges to examine relevant aspects (Dibbern et al., 2004). Findings repeatedly align with Dyer and Singh's (1998) emphasis on informal governance for value-creation initiatives. Informal mechanisms can involve clan control, which relies on an implicit system of shared values that promote desirable behavior, or self-control, which encourages self-monitoring (Kirsch, 1997; Wiener et al., 2016).

In the reviewed literature, it is widely understood that the outsourcing engagement must not be treated as an arm's-length, tactical relationship (Barua and Mani, 2014; Lahiri and Kedia, 2009; Susarla and Mukhopadhyay, 2019), but rather as a partnership (Kotlarsky et al., 2015; Levina and Su, 2008; Weeks and Feeny, 2008). Certain characteristics are generally associated with partnerships, including shared interests (Bui et al., 2019; Frydlinger et al., 2019; Weeks and Feeny, 2008), high levels of trust (Kedia and Lahiri, 2007; Søderberg et al., 2013; Weeks and Feeny, 2008) and transparency through shared expectations (Henke Jr. and Zhang, 2010; Søderberg et al., 2013), irreversible investments in provider-specific technologies (Susarla and Mukhopadhyay, 2019), a strong identification with the project among participants (Søderberg et al., 2013), mid-level and corporate executive involvement (Handley and Benton, 2009; Miranda and Kavan, 2005; Weeks and Feeny, 2008), shared routines (Argyres et al., 2007), and enabling control styles that promote cooperation (Wiener et al., 2019, 2016).

Prior work commonly examines how these partnership elements and informal control mechanisms interact with configurational and contractual aspects when attempting to innovate in outsourcing. In accordance with Poppo and Zenger (2002), recent research demonstrates that relational norms can effectively complement contractual safeguards to motivate continuous cooperation and increase the likelihood of realizing innovation through outsourcing (Susarla and Mukhopadhyay, 2019). Weeks and Feeny (2008) find that in the most successful strategic innovation through outsourcing initiatives, the parties rely on a trust-but-verify approach, which builds on tightly maintained service levels.

In contrast, other studies propose that complex tasks like those associated with strategic innovation can be completed more successfully when relationship-building elements are clearly prioritized over contractual specifications (Langer and Mani, 2018). The latter should be relaxed, while autonomy, incentives, and trust should be strengthened (Bui et al., 2019; Vitasek and Manrodt, 2012). This in turn should encourage the provider to step outside of formal boundaries and explore promising ideas that have not been defined in advance (Aubert et al., 2015). The lack of contractual safeguards, however, implies that such approaches are highly prone to relational risks.

### *Arrangement phase: Conclusion*

Findings included in the arrangement phase appear both diverse and inconsistent. We categorized prior insights into five themes: relational risk portfolios, outsourcing configurations, pricing models, contract design, and relationship management. The first theme, risk portfolios, illustrates common relational risks that are considerably amplified when leveraging outsourcing for strategic innovation. Configurations vary greatly in terms of supply base breadth. Slim and broad supply bases each come with their own set of advantages and disadvantages. The literature related to pricing models presents a slightly more consistent picture, with flexibly priced contracts seemingly superior to fixed-price contracts. Contract design has inspired a largely dichotomized debate between scholars arguing that more complete contracts can facilitate innovation, while others suggest that strategic innovation, which is difficult to specify (Kotlarsky et al., 2015), requires loose contractual regimes. Lastly, while it is widely agreed that partnerships are vital for innovation through outsourcing, it remains unclear to what extent the parties should balance formal with relational mechanisms. Table 4 summarizes our findings.

**Table 4: Overview of key findings related to the "Arrangements" phase**

| Main theme | Key findings |
|---|---|
| Relational risk portfolio | Providers may hide information to win innovation through outsourcing contracts, resulting in adverse selection problems. After entering engagements, clients are exposed to providers who hide changes in their behavior. They may more specifically secretly shirk their innovation-related responsibilities. Both parties are further at risk of experiencing hold-ups and consequent retaliatory behavior. Providers may also share commercially sensitive domain knowledge of the client with their other customers. |
| Outsourcing configurations | Engaging with one or a few providers (slim supply bases) or multiple providers (broad supply bases) each comes with its own advantages and shortcomings, especially in terms of access to a diversity of innovative ideas, dependency on service provider(s) and coordination issues. Slim and broad supply bases can be combined using provider ranking systems. |
| Pricing models | Fixed price contracts are generally viewed as disadvantageous when pursuing strategic innovations in outsourcing engagements. Flexible pricing models are widely found to be the more favorable alternative. Equity-based contracts entail high set-up costs, but apparently deliver consistently good results. Performance-based contracts can similarly stimulate strategic innovations, but require reinforcing conditions. |
| Contract design | Research is inconsistent regarding contract completeness. In general, the topic is dominated by a dichotomous debate between more complete contracts that may facilitate strategic innovation efforts with targeted controls like joint innovation boards, and more incomplete contracts that provide greater autonomy and can propel creativity. |
| Relationship management | Scholars widely agree that outsourcing engagements should not be approached as transactional relationships, but rather as business partnerships when pursuing strategic innovations. Yet, in view of the inconsistencies pertaining to the degree of formalization, it remains unclear to what extent the partners should rely on informal governance. |

### Phase 3: Generating strategic innovations through outsourcing

The third phase of the integrative framework encompasses researched aspects relating to the joint generation of strategic innovations. Arguably, innovation depends increasingly on the ability to utilize new knowledge produced elsewhere and to combine this with already available knowledge (Hoecht and Trott, 2006). In an outsourcing context, the client usually possesses deep knowledge specific to its domain, while the provider possesses deep technological knowledge (Chatterjee, 2017; Oshri et al., 2018). These knowledge bases need to be synthesized for the generation of strategic innovations.

Our findings are categorized into two themes – knowledge combinations and architectural coordination (see Figure 4). The former deals with knowledge flows within the project environment, while the latter comprises organizational and technological architectures that may facilitate these knowledge flows.

**Generation**
*Collaborative development activities and strategic innovation integration*

| Main themes | Characteristics specific to outsourcing | Key theoretical perspectives |
|---|---|---|
| Knowledge combinations | Innovation-related knowledge exchanges span across organizational boundaries | RBV, KBV, learning, familiarity, boundary spanning, practice and cultural perspective |
| Supporting architectural structures | The client's IT function is the main subject of outsourcing decisions and is typically closely linked to many other business functions of the client | RBV, KBV, learning, boundary spanning |

**Figure 4. Notable characteristics and theoretical perspectives in the "Generation" phase**

### *Knowledge combinations*

Theoretical foundations of prior research relevant in the generation phase range from the RBV to the KBV, and related perspectives like absorptive capacity and learning, problem-solving (Nickerson and Zenger, 2004), cultural differences (Hofstede, 2003), familiarity (Herrera and Blanco, 2011), practice theory (Bourdieu and Wacquant, 1992), and boundary spanning (Carlile, 2002). Two project-related issues recur in the literature, namely the transfer of tacit domain knowledge and the role of acculturation.

A core issue related to the generation of strategic innovations through the combination of knowledge bases is the transfer of tacit domain knowledge (Chatterjee, 2017; Roy and Sivakumar, 2012; Weigelt and Sarkar, 2012). Consistent with the KBV, tacit knowledge requires context-specific understanding to make sense (Weigelt, 2009). The provider primarily absorbs client-specific domain knowledge through repeat interactions with the client (Oshri et al., 2018), particularly via learning by doing and trial and error (Chatterjee, 2017). Forging deep social ties (Miranda and Kavan, 2005) and developing a shared language (Barua and Mani, 2014) is vital, and accentuates the importance of arranging the engagement as a partnership.

Concerning acculturation, the research consistently notes that cultural distances need to be minimized to enable innovation in outsourcing (Chen and Lin, 2019; Lacity and Willcocks, 2013). Such findings align with KBV arguments when applied to inter-organizational networks, which emphasize the importance of building shared identities in a network (Dyer and Nobeoka, 2000). Cultural differences manifest in form of differing employee values and norms, attitudes towards technology, customers, interpersonal contact and interaction, and role perceptions (Kedia and Lahiri, 2007).

### *Supporting architectural structures*

Outside of the project environment, we identified two areas in the broader organizational context that are closely connected to the generation of strategic innovations through outsourcing. First, focusing on the organizational architecture of the client, we present prior findings that suggest retaining an IT function. Second, research indicates that the client's technological architecture needs to incorporate common standards in order to seamlessly integrate jointly generated strategic innovations.

Whereas domain knowledge represents "know-why", technological knowledge refers to the "know-how" necessary to customize vanilla solutions to meet the specific business objectives of the client (Chatterjee, 2017). There is broad agreement that the client must retain its internal IT function to absorb technological knowledge (Weigelt and Sarkar, 2012). Prior evidence shows that an internal IT function that is too underdeveloped can inhibit effective communication with the provider (Weeks and Feeny, 2008). Then again, a retained IT function that is too strong can be a source of conflict (Miozzo and Grimshaw, 2005), or lead to duplicated efforts (Weigelt, 2013).

To integrate strategic innovations successfully, the literature suggests that technological architectures need to be in place which feature common application and data standards (Su et al., 2015; Sumo et al., 2016). Scholars also find that successfully integrating new IT solutions does not guarantee their use (Chatterjee, 2017; Hong and Zhu, 2006; Weigelt, 2013, 2009). Bypassing the potentially slow internal development process may not only result in an innovation that is poorly aligned with the client's other business processes and operational capabilities (Hong and Zhu, 2006; Weigelt, 2013), but may also lead to a strong political bias, with the innovation being regarded as a "foreign" solution (Hong and Zhu, 2006; Weigelt, 2009).

### *Generation phase: Conclusion*

This phase captures notable thematic patterns related to the joint generation of strategic innovations through outsourcing (see Table 5). Usually, generation efforts presuppose an in-depth familiarization between the client and provider. The provider's general IT products can thereby be infused with client-specific domain knowledge that enables high degrees of customization. Prior research also indicates that close collaboration for strategic innovation generation is vital from the early exploration stages onwards. Reducing cultural differences similarly facilitates innovation activities. Perhaps the most important insight in this phase, and one which rewrites the traditional rulebook for outsourcing, is the indispensability of retaining IT function at the client when pursuing strategic innovation.

**Table 5: Overview of key findings related to the "Generation" phase**

| Main theme | Key findings |
|---|---|
| Knowledge combinations | Within the outsourcing project, the enablement of tacit domain knowledge flows is crucial to enable new knowledge combinations for customized innovations. Generation efforts may further be promoted by bilateral, rather than one-sided contributions, by the close involvement of the service provider from the early stages of high-level design up to the late stages of innovation integration, and by reduced cultural distances. |
| Supporting architectural structures | Supporting structures outside of a specific outsourcing project may catalyze the generation of strategic innovations. Research is largely in agreement that the client firm must possess deep technological knowledge to effectively collaborate with its service provider. Accordingly, the internal IT function needs to be retained or rebuilt. Furthermore, a compatible technological architecture with boundary conditions that are clearly visible to all parties involved has been suggested to be beneficial. |

## Phase 4: Outcomes of strategic innovations through outsourcing

The fourth phase in the framework captures scholarly assessments of benefits and challenges derived from the implementation of strategic innovations. Prior insights are divided into five themes (see Figure 5). The first theme introduces researched business benefits from the client and provider perspective. Central issues with measuring created benefits are discussed next, followed by appropriability mechanisms that may reduce replication risks from knowledge leakages, hollowing out concerns, and arguments about the uniqueness of strategic innovations.

**Outcomes**
*Value creation and value capture aspects in strategic innovation*

| Main themes | Characteristics specific to outsourcing | Key theoretical perspectives |
|---|---|---|
| Realizable business advantages | Strategic innovation enables substantial back-end and front-end improvements for clients and project scope expansions for providers | IT-based innovation typologies |
| Value creation measurement | The imperfect observability of provider efforts can hurt innovation outcomes | Agency theory |
| Knowledge leakage | Providers as external firms are expected to offer innovative solutions for all their clients | KBV, intellectual property and appropriability strategy perspectives |
| Hollowing out concerns | The erosion of IT resources over the long-term is a well-recognized outsourcing concern | RBV |
| Output uniqueness | IT-based solutions tend to be customized modifications of a provider's state-of-the-art offerings | RBV |

**Figure 5. Notable characteristics and theoretical perspectives in the "Outcomes" phase**

### *Realizable business advantages*

The literature shows that outsourcing can be successfully leveraged for a variety of custom solutions that have been linked to different innovation typologies, including by Weeks and Feeny (2008) and Swanson (1994). Such custom solutions may enable business advantages in the form of substantial back-end improvements, consequently improving the client's operating efficiency, business process effectiveness, and strategic performance (Lacity and Willcocks, 2013). Clients may also realize business advantages in the form of extensive customer-facing enhancements that complement, adapt, or extend the usage of their offerings (Susarla and Mukhopadhyay, 2019).

There is a paucity of research assessing realizable business advantages from the provider perspective. Some studies nonetheless argue that introduced innovations act as a gateway for securing additional future business opportunities with the same, but more satisfied client (Gopalakrishnan and Zhang, 2017; Henke Jr. and Zhang, 2010; Oshri et al., 2015). The client may be compelled to increase the scope of transferred value-creating activities in order to encourage the provider to deliver more innovations (Gopalakrishnan and Zhang, 2017). Strategic innovation may thus enable business advantages in the form of higher project revenues for the provider (Oshri et al., 2015), and possibilities for leveraging client-specific investments in follow-on innovation initiatives (Linder, 2004; Susarla et al., 2010).

### *Value creation measurements*

Measuring the performance outcomes enabled by strategic innovations is challenging as they tend to manifest in multiple performance dimensions (Susarla et al., 2010). Oshri et al. (2018), for instance, measure innovation outcomes along the dimensions of innovation quality, innovation frequency, cost savings and service improvements. Some dimensions, however, are qualitative, which complicates the verifiability of certain aspects of performance (Langer and Mani, 2018; Susarla et al., 2010). The literature nevertheless largely agrees that tracking measurable outcomes is paramount (Linder et al., 2003). Clients that do not actively track measurable outcomes expose themselves to moral hazards, which may result in low-quality outputs and underwhelming business performance outcomes (Roy and Sivakumar, 2012; Shi, 2007).

### *Knowledge leakage risks*

More knowledgeable providers can more effectively engage in innovation efforts on behalf of the client (Chatterjee, 2017; Oshri et al., 2018). However, in the tradition of the KBV, knowledge leakage risks increase as the client shares more valuable domain knowledge with the provider (Dyer and Nobeoka, 2000; Hoecht and Trott, 2006). Providers are expected to deliver custom solutions for all their customers, and therefore unlikely treat domain knowledge as exclusive to a specific client (Hoecht and Trott, 2006). Clients in contrast want to prevent the strategic innovation and associated commercially sensitive knowledge being sold on to their competitors (Miozzo et al., 2016). This mirrors the knowledge poaching risk (Aron et al., 2005; Handley and Benton, 2009).

The intellectual property (IP) management and appropriability strategy literature form the theoretical basis of most related research. Appropriability mechanisms are widely labelled as formal, including patents, copyrights, design rights or trademarks, and informal, including lead-time advantages, product complexity, complementary assets and secrecy (Desyllas et al., 2018; Miozzo et al., 2016). In an outsourcing context, Leiponen (2008) finds that providers focusing on customized solutions put little emphasis on formal IP rights. In fact, they are often willing to transfer these rights to the client. Desyllas et al. (2018) similarly find that providers competing on the basis of customized solutions are less disposed to rely on formal appropriability mechanisms than their cost-oriented rivals; their innovations are better protected with informal appropriability mechanisms. This is echoed in Miozzo et al.'s (2016) results, which show that modest levels of emphasis on formal appropriability mechanisms are best suited to preventing unintended knowledge leakages.

### *Hollowing out of client's internal IT resources concerns*

Another potential negative outcome brought forward in prior research is the gradual hollowing out of corporations (Mukherjee et al., 2013; Weigelt, 2009). This is a well-documented issue in the IS sourcing literature and rooted in the principles of the RBV, which postulate that valuable resources are scarce, difficult to imitate and substitute, and evolve within the firm (Barney, 1991). When engaging in outsourcing, the client surrenders the development of its resources to the provider (Shi, 2007). Consequently, the client's internal resources depreciate over time, which compromises its ability to exploit future business opportunities (Miozzo and Grimshaw, 2005; Weigelt, 2009).

To what extent hollowing out concerns take effect in the strategic innovation through outsourcing context remains unclear. In essence, the consequences seem to be similar to traditional outsourcing outcomes. Overreliance discourages the client from developing its internal technological knowledge base (Lee and Kim, 2010; Manning et al., 2018). As the client gradually loses its ability to detect and exploit new IT-enabled opportunities, it becomes less innovative (Hoecht and Trott, 2006) and more dependent on the provider to show leadership in innovation (Lee and Kim, 2010).

### *Uniqueness of strategic innovations*

Lastly, the uniqueness of the strategic innovations achieved through outsourcing has been called into question. Here again, central tenets of the RBV largely form the theoretical foundation. Some studies suggest that providers offer seemingly bespoke solutions, which, however, tend to be standardized based on industry best practices (Shi, 2007; Weigelt and Sarkar, 2012). Customized IT products may be targeted at a vertical segment or may cut across segments, but are rarely specific to individual clients (Arora et al., 2001).

These views contrast with studies that emphasize the uniqueness of localized innovations (Avgerou, 2008). Greater customization to idiosyncratic business needs provides the client with a highly firm-specific innovation (Kedia and Lahiri, 2007; Lema et al., 2015) that its rivals may find difficult to replicate (Qu et al., 2010). Such innovations based on unique knowledge combinations can lead to competitive advantages (Mani et al., 2010). While the provider may salvage its core, which tends to be a replicable service (Arora et al., 2001; Desyllas et al., 2018), the provider will find it difficult to redeploy the innovation with identical content to its other clients (Desyllas et al., 2018; Mani and Barua, 2015). Ultimately, the question of innovation uniqueness and enabled competitive advantages seems to boil down to the degree to which the strategic innovation is contextualized to the client.

### *Outcomes phase: Conclusion*

This phase synthesizes scholarly assessments of outcomes associated with strategic innovation pursued in the context of outsourcing. Five main themes were developed to compare and contrast prior findings. The first theme includes realizable business advantages from the client and provider perspective. In the second theme, value creation measurement issues are presented, largely stemming from non-contractible investment returns. The third theme introduces knowledge leakage risks that may be effectively mitigated with informal appropriability mechanisms. The fourth theme highlights that hollowing out concerns in relation to client IT resources are unclear when the client retains its IT function. The fifth theme shows diverging views on innovation uniqueness, which may be reconciled by recognizing the presence of different degrees of strategic innovation customization. Table 6 provides an overview of key findings.

**Table 6: Overview of key findings related to the "Outcomes" phase**

| Main theme | Key findings |
|---|---|
| Realizable business advantages | Custom IT products can enable improved back-end and front-end business operations for the client firm, and new business opportunities for the service provider when extending the relationship. |
| Measuring value creation | Multiple performance dimensions, some of a qualitative nature, complicate measurement of created business value for the client firm. |
| Knowledge leakage risks | Replication risks stemming from knowledge leakages can be effectively mitigated with informal appropriability mechanisms. |
| Hollowing out of client's IT resources | The traditional outsourcing risk of atrophying internal knowledge remains active when client firms rely heavily on service providers for innovation. |
| Strategic innovation uniqueness | Strategic innovations only enable competitive advantages when they are sufficiently contextualized. |

## Discussion

Our review maps out a strategic innovation through outsourcing research landscape that is becoming ever more complex. In parallel with this growing complexity however, we note uneven scholarly interest in specific phases of our framework. Most of the research to date appears to concentrate on the first two phases of our integrative framework, Antecedents and Arrangement, while the Generation and Outcomes phases only receive scant attention. Furthermore, the majority of reviewed literature focuses on a client perspective (58 papers), while the provider (20 papers), and bilateral perspectives (17 papers) are less represented.

Moreover, there is a notable lack of concepts incorporated from the innovation literature. This obscures challenges that may emerge specifically during the discovery, development, and integration of collaborative strategic innovation initiatives. More recent advancements in the digital innovation literature (Henfridsson et al., 2018; Nambisan et al., 2017; Yoo et al., 2012, 2010) introduce concepts that may better fit the contemporary zeitgeist in the outsourcing industry, such as digital technology editability (Yoo et al., 2010), non-linear digital innovation processes (Nambisan et al., 2017), and design and use recombination (Henfridsson et al., 2018). Nambisan et al. (2017) suggest that "the transition from innovation to digital innovation comes as a golden opportunity" (p. 224) that must be seized by IS scholars. We argue that this line of thought similarly applies to the strategic innovation in outsourcing context.

We also note inconsistencies in the way innovation is conceptualized in the outsourcing context. Appendix VI includes a column showing an overview of study-specific concepts, descriptions, definitions, or empirical examples that fit our conceptualization of strategic innovation. There seem to be differences on what qualifies as strategic innovation, depending on the adopted perspective. From the client perspective, strategic innovation accommodates radically new IT-based products and services that overturn established principles of operation. From the provider perspective, strategic innovation involves reconfiguring their "best practice" solutions by linking key components together in new ways that fit a client's specific business context (Desyllas et al., 2018). In this regard, additional research that examines how strategic innovations differ from the client and provider perspectives may not only shed more light on the concept itself, but also bring clarity to the often encountered disagreement between practitioners from the client and provider side as to what constitutes a strategic innovation (Weeks and Feeny, 2008).

Finally, research that scrutinizes failed innovation initiatives remains largely absent from the body of knowledge. In a practitioner-informing discipline like IS sourcing (Lacity et al., 2009), we consider this a critical problem. Evidently, failed innovation initiatives are not uncommon in practice (Weeks and Feeny, 2008; Whitley and Willcocks, 2011). More research in this vein may, for instance, uncover different states of digital readiness (Gfrerer et al., 2021) for strategic innovation efforts, or different reasons for failure and subsequent conflicts, and effective conflict resolution strategies (Lacity and Willcocks, 2017).

*How leveraging outsourcing for strategic innovations differs from cost savings objectives*

There appear to be several ways in which the literature focusing on innovation in an outsourcing context differs from traditional research on outsourcing engagements, which is typically (but not exclusively) driven by cost savings motives. Building on Aubert et al. (2015), who distinguish between an innovation-oriented and contractual view of outsourcing, and Dibbern et al.'s (2004) review of IS outsourcing literature published between 1988 and 2000, we present notable differences and overlaps in each of our explored themes in Table 7.

Principally, we observe considerable differences in complexity, not just at the task level, but also in terms of engagement arrangements and organizational architecture adjustments. Nevertheless, there are certain areas, including the relational risk portfolio and outsourcing configurations, that do not appear to have changed drastically with a shift from cost savings to strategic innovation objectives. It must therefore be emphasized that strategic innovation through outsourcing does not constitute an entirely new phenomenon, but rather an evolution of traditional cost-oriented outsourcing practices that builds on a legacy of pioneering work in the IS sourcing literature.

Table 7 suggests a noticeable shift in theoretical foundations. This is visible in the diversity of theoretical foundations employed in innovation-oriented outsourcing research (see Appendix IV for details), and contrasts with the high concentration of TCE-based studies that form the core of cost-oriented outsourcing engagement research (Dibbern et al., 2004). Major differences can also be recognized in the *task attributes* and *organizational considerations* themes of the Antecedents phase. Regarding the former, the traditional task environment involves relatively simple and standardized generic business services that are amenable to outsourcing, while innovation-related tasks are of an explorative nature and complex (Aubert et al., 2015). Concerning the latter, the cost savings rationale is often based on early views of IT as a utility (Lacity and Hirschheim, 1993), whereas IT is associated with vast strategic business outcome potential in innovation-oriented outsourcing (Weeks and Feeny, 2008). An implicit consequence of these differences is a stronger link between outsourcing and the client's overall business strategy.

Notable differences are further evident in themes associated with the Arrangement phase. In *pricing strategies*, the traditional IS sourcing body of research commonly examines engagements governed by fixed price or time and materials contracts (Dibbern et al., 2004), which clearly remain widespread in practice (Oshri et al., 2015). Based on the reviewed literature however, their adequacy for strategic innovation objectives appears questionable (Bui et al., 2019; Oshri et al., 2015).

Evolving relationship styles also warrant more attention. Cost-oriented engagements are typically governed by arm's-length relationships (Miranda and Kavan, 2005), characterized by a short-term focus and low commitment (Barua and Mani, 2014). In contrast, partnership-styled relationships are usually regarded as indispensable for innovation-oriented engagements (Kotlarsky et al., 2015). Key assumptions of common theoretical perspectives need to be adapted accordingly. Following Wiener et al. (2019), agency theory for instance may be well attuned to studying transactional IS projects because it assumes a short-term orientation. However, the theory is a poor fit for partnership-based relationships. Instead stewardship theory (Donaldson and Davis, 1991), which assumes a long-term orientation, may be a more suitable alternative to study such relationships.

Lastly, we want to point out two other notable changes, one in the *architectural coordination* theme and the other in the *knowledge leakage risks* theme. In the former theme, traditional outsourcing guidelines based on core competency perspectives propose that the IT function needs to be outsourced so the client can focus on its core business (Grover et al., 1996). These guidelines have been gradually superseded by the view of the IT function as a core part of the client that facilitates its competitive activities (Baldwin et al., 2001; Peppard, 2018). To effectively communicate with the provider when developing strategic innovations, research suggests that the IT function needs to be retained by the client (Weeks and Feeny, 2008). However, it remains unclear what a communication-encouraging social context looks like. In the latter theme, unintentional knowledge leakages are portrayed as a much greater concern in innovation-oriented engagements than in their cost-oriented counterparts due to extensive transfer of domain knowledge.

**Table 7: Juxtaposition of cost-oriented and innovation-oriented outsourcing engagements**

| | Theme | Cost-oriented outsourcing characteristics | Strategic innovation-oriented outsourcing characteristics |
|---|---|---|---|
| Antecedents | Theoretical underpinnings | Agency theory, RBV, TCE (Aubert et al., 2015, 2004; Dibbern et al., 2004) | Distributed innovation perspectives, knowledge management, organizational design (Aubert et al., |
| | | | 2015), relationship management (Chou et al., 2015) |
| | Task attributes | Simple, easy to measure and standardized (Aubert et al., 2004) | Highly uncertain (Aubert et al., 2015), unstructured (difficult to codify) and complex (Weigelt and Sarkar, 2012) |
| | Organizational considerations | Pursuit of cost savings is driven by the view of IT as a utility that can be outsourced (Lacity and Hirschheim, 1993) | Successful prior collaborations and specialized resources can promote IT-enabled business development (Desyllas et al., 2018; Weeks and Feeny, 2008) |
| | Environmental conditions | Bandwagon effects drive outsourcing decision (Lacity and Hirschheim, 1993; Loh and Venkatraman, 1992a) | Network-based competition (Van de Ven, 2005) and growing need for service differentiation (Arora et al., 2001) as drivers |
| Arrangement | Relational risk portfolio | Adverse selection, moral hazard, hold-up problem (Loh, 1994) | Same relational risks, but amplified, plus poaching (Aron et al., 2005; Handley and Benton, 2009) |
| | Outsourcing configurations | Single or multi-sourcing (Gallivan and Oh, 1999) | Single or multi-sourcing (Su and Levina, 2011; Weeks and Feeny, 2008) |
| | Pricing strategy | Typically fixed-price or time and materials contracts (Currie, 1996; Gopal et al., 2003) | Flexible pricing is essential (Bui et al., 2019; Oshri et al., 2015) |
| | Degree of formalization | Emphasis on tighter, complete contracts (Aubert et al., 2015; Currie, 1996) | Looser contracts or more innovation-oriented terms may facilitate joint innovation efforts (Aubert et al., 2015; Weeks and Feeny, 2008) |
| | Relationship management | Arm's-length/transactional relationship style (Lee et al., 2004) | Partnership relationship style (Kotlarsky et al., 2015) |
| Generation | Knowledge combination | Mainly technological resource exchanges limited to IT functions (Grover et al., 1994) | Intensive domain and technological knowledge exchanges (Chatterjee, 2017) |
| | Architectural coordination | Downsized or fully outsourced IT function (Dibbern et al., 2004) | Retained IT function (Weeks and Feeny, 2008) |
| Outcome | Realized business advantages | Immediate economic (Gallivan and Oh, 1999) and operational (Dibbern et al., 2004) efficiencies; created value is independent of business strategies (Venkatraman, 1997) | Substantially improved overall business performance (Kotlarsky et al., 2015; Weeks and Feeny, 2008) |
| | Outcome measurement | Efficiency metrics, such as cost per millions of instructions per second (Venkatraman, 1997) | Variety of metrics needed to measure multi-dimensional outcomes (Linder et al., 2003; Susarla et al., 2010) |
| | Knowledge leakage risks | Security concerns mainly involve physical IT assets, software and data (Fink, 1994) | Commercially sensitive domain knowledge may be leaked (Hoecht and Trott, 2006) |
| | Hollowing out of client's IT resources concerns | Ability of the client firm to compete with IT is adversely affected over time (Willcocks et al., 1995) | Yet unclear effects when the IT function is retained |
| | Output uniqueness | Provided services are generic, such as daily processing runs or back-up procedures (Grover et al., 1994) | Highly customized innovations may be unique (Lacity and Willcocks, 2013) and enable competitive advantages (Kotlarsky et al., 2015) |

*Recommended research directions*

In this section, we introduce five research directions (see Figure 6). The first direction reflects the changing nature of innovation in an outsourcing context with respect to the increasing prevalence of digital innovation and digital transformation. These developments are expected to have considerable implications for all phases in our framework. We also identify an underexplored area in each of the four phases that may particularly benefit from research efforts based on a closer orchestration of recent advances in the digital innovation and strategic innovation through outsourcing literatures.

Specifically, we recommend: (i) studying imitation as a driver of strategic innovation; (ii) understanding equal and unequal contributions of resources by the client and provider; (iii) examining microfoundations and mechanisms of knowledge combination; and (iv) considering how to manage potential knowledge leakages while dealing with IP concerns.

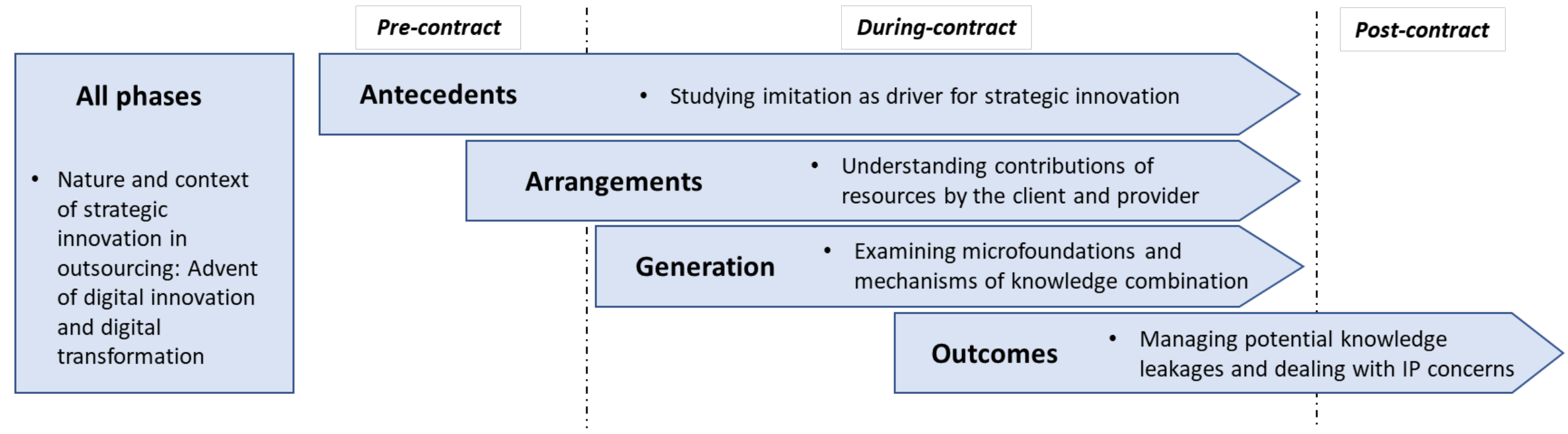

**Figure 6. Five recommended research directions**

***Nature and context of strategic innovation in outsourcing: Advent of digital innovation and digital transformation***

Current digitalization trends, such as wearables, cloud computing, internet-of-things, mobile, social media and data analytics (Lokuge et al., 2019), facilitate (re)combinations of existing products and services to generate new forms of digital offerings. Related research in an outsourcing context, however, is still at a nascent stage (Dibbern and Hirschheim, 2020). At the same time, the outsourcing industry is increasingly focused on offerings based on cutting-edge technology, such as cloud-based services, data analytics and robotic process automation. We see a clear convergence of digital innovation and outsourcing practice. We therefore suggest that future research on strategic innovation through outsourcing considers relevant insights from the growing body of IS literature on digital innovation. Digitalization is leading to re-examination of innovation management theories, by questioning fundamental assumptions regarding the boundaries of innovation (Nambisan et al., 2017). Digital innovation is more dynamic, compared to innovation in the realm of physical products and services. As a result, digital innovation defies the traditional boundaries of the innovation process and the organization, making the process more open to external parties. Properties of digital artifacts (Kallinikos et al., 2013) enable open-ended value creation through the recombination of digital components (Henfridsson et al., 2018). This in turn, makes the external environment of an organization an important source of innovation (Kohli and Melville, 2019) rather than an area of hostile competition.

We believe there is a great opportunity for sourcing scholars to bring a new perspective to digital innovation discourse, and vice versa. There is potential in exploring new innovation management practices of client and provider firms that are encouraging a change in attitude towards both digital innovation processes and outsourcing engagements. One could explore whether shifting from managing outsourcing through contracts to some form of provider orchestration akin to innovation ecosystems might increase the innovative potential of outsourcing engagements. The motivations of client and provider firms to engage in digital innovation projects is another aspect warranting further research.

Furthermore, closely related is the growing trend towards digital transformation (Vial, 2019; Wessel et al., 2020). More and more firms are contracting providers to help them on their digital transformation journey. Such journeys typically involve attempts to significantly re-design and digitize business processes, or establish new digital revenue. They may thus comprise waves of digital innovations that involve service providers and result in broader processes of transformative change (Holmström, 2018; Vial, 2019). How such a comprehensive change, which fundamentally alters the fabric of the client firm (Vial, 2019), can be achieved remains underexplored in IS sourcing research. We therefore see significant potential for outsourcing scholars to carry out in-depth studies of various aspects of strategic innovation in an outsourcing context, by focusing on digital transformation projects that involve clients and providers.

***Antecedent phase: Studying imitation as a driver for strategic innovation***

This research direction builds on the finding that outsourcing relationships are increasingly leveraged for innovations that support a firm's overall business strategy. Yet, there appears to be little research that examines this outsourcing–business strategy link in more depth. First advances have been made by Ruckman et al. (2015) from a provider perspective. They show that imitation is a useful business strategy to quickly expand and consolidate service lines in response to client demands, and to develop follow-on innovations rather than "knock-offs" in order to create more diverse offerings. This points to imitation as an understudied but potentially highly influential driver for the strategic innovation through outsourcing decision from a client perspective.

In the innovation literature, imitation usually refers to a business strategy that involves copying the innovator or stronger competitors to achieve business growth and increase profits (Levitt, 1966). Coupled with our more general recommendation to study strategic innovation through outsourcing in light of emerging digital technologies, we suggest that more research on imitation-based digital business strategies (Bharadwaj et al., 2013) may offer valuable insights for contemporary outsourcing literature.

***Arrangement phase: Understanding contributions of resources by the client and provider***

Our review also reveals that equity-based arrangements are especially conducive to innovation. Such arrangements, however, continue to remain poorly understood compared to fixed-price or time and materials contracts (Currie, 1996; Gopal et al., 2003). We propose examining how client and provider resource contributions in equity-based arrangements influence joint strategic innovation initiatives.

Oshri et al. (2015) for example, report that joint venture contracts in particular improve relationship quality and thereby deflate the risk of opportunism. Building on these insights, future research could not only examine how equity-based arrangements affect the many elements of partnerships, such as trust and shared interests, but also explore under which circumstances clients and providers will be more or less encouraged to develop and maintain equally contributed, relationship-specific resources.

Furthermore, connected to the advent of digital technologies, it may be worthwhile studying the role of relationship-specific digital resources (Henfridsson et al., 2018), both of a technological nature and intellectual nature. Attitudes towards the contribution of such resources may change over time. It could therefore be useful to examine (digital) resource contributions in equity-based arrangements with a dynamic perspective. Different stages of innovation development, such ideation or prototyping (Kotlarsky et al. 2016), may serve as reference points for studying mid-initiative resource contribution decisions from the client and provider perspective.

***Generation phase: Examining microfoundations and mechanisms of knowledge combination***

With this recommended research direction, we seek to respond to the paucity of in-depth research on knowledge combination facilitation mechanisms. The reviewed literature clearly demonstrates that bilateral knowledge flows are critical for the generation of strategic innovations (Chatterjee, 2017; Oshri et al., 2018), making this a worthwhile area to investigate.

Traditional outsourcing literature provides some insights that can be built on. Quinn and Hilmer (1994) for instance suggest that close personal relationships between operating-level personnel can be promoted when provider specialists physically relocate to the client’s premises for development projects. More recently, Oshri et al. (2018) studied the role of advisors, finding that their contribution to innovation is conditional on an already established shared understanding between the client and provider, as well as on the provider's knowledge of the client.

We present these insights in particular because they imply the importance of examining attitudes and behaviors of organizational members. This promises a deeper understanding of the mechanisms that promote the transfer of domain/technological knowledge and facilitate new knowledge combination activities. Insights from the recent IS literature on digital innovation (e.g., Svahn and Mathiassen, 2017) could provide further theoretical and practical insights on the generation of digital innovation through outsourcing.

***Outcome phase: Managing potential knowledge leakages and dealing with IP concerns***

This recommended direction calls for research to address the challenge of managing knowledge leakages and dealing with IP concerns, when innovations are concerned with digital products and services. Knowledge leakages appear to be less of an issue when limited domain knowledge is required to deliver generic business services, which explains why they have been of little relevance in studies examining traditional outsourcing engagements (Dibbern et al., 2004). They are however a major risk when the client shares commercially sensitive information with its provider (Hoecht and Trott, 2006). While there is growing evidence that informal appropriability mechanisms and trust can counteract unintentional knowledge leakages, there is still much to explore.

Strategic management research on supply relationships shows that clients like Toyota can afford knowledge leakages in their supplier networks, as long as they move faster than their rivals in deriving learning advantages from suppliers (Dyer and Nobeoka, 2000). Future research could examine under which circumstances clients can afford knowledge leakages when collaborating with providers on strategic innovation in an IS outsourcing context, as well as how to protect IP associated with strategic innovations developed jointly by clients and providers.

## *Contributions and Limitations*

This review offers three major contributions. The first contribution is the consolidation of a large body of knowledge into an integrative framework. The framework organizes insights from 95 reviewed papers along four higher-order phases that emerged from our analysis of the literature, starting with the decision to leverage outsourcing strategic innovations, followed by outsourcing arrangements, the generation of strategic innovations, and finally their outcomes. Our findings show that prior research has utilized various theoretical lenses to study a range of aspects related to the strategic innovation through outsourcing phenomenon. Knowledge management and relationship management theories are especially common, while there is a surprising lack of innovation theories.

This helps explain why research to date largely focuses on optimal arrangement configurations, but neglects issues associated with collaborative generation efforts.

Our second contribution aligns with our aim to discuss how innovation-oriented outsourcing compares with cost-oriented engagements. We draw on early IS sourcing works and related reviews for this juxtaposition. We find that strategic innovation in outsourcing presents an evolution rather than a revolution of outsourcing practices and that prior findings are consciously built on and extended. Our third contribution is the presentation of five recommended research directions. They should help scholars locate areas that are worthwhile for closer examination and produce findings of high theoretical and practical relevance, especially in view of emerging digital technologies.

This review is subject to some limitations. First, our formal quality appraisals limit the article sample to a high-quality, albeit limited set of papers. Here, we acknowledge the abundance of relevant studies in excluded outlets. Many papers presented in conference proceedings also feature insights that may be of great value to the research stream. Second, our review scope isolates the paper sample to our conceptualization of strategic innovation through IS outsourcing. As discussed, innovation in an outsourcing context may evoke broader associations, such as innovation in a contract manufacturing context (Cabigiosu et al., 2013; Preeker and De Giovanni, 2018).

## Acknowledgements

We thank the Senior Editor and the three anonymous reviewers whose comments and guidance helped us to greatly improve the article. We further want to thank Prof. Jan Mendling and his research group for their feedback on an earlier version of this paper presented at the Business Process Mining virtual seminar 2020 at the Vienna University of Economics and Business. Lastly, we appreciate support in form of a research grant from the ACM Special Interest Group on Management Information Systems (SIGMIS).